\documentclass[useAMS,usenatbib,a4paper]{mn2e}

\usepackage{epsfig,bm}
\usepackage{amsbsy,amssymb,amsmath}
\usepackage{hyperref}
\usepackage{graphicx}
\usepackage{dcolumn}
\usepackage{ctable}
\usepackage{natbib}
\usepackage{float}
\usepackage{color}

\newcommand{\be}{\begin{equation}}
\newcommand{\ee}{\end{equation}}
\newcommand{\bea}{\begin{eqnarray}}
\newcommand{\eea}{\end{eqnarray}}

\newcommand{\bn}{\mathbf{n}}
\newcommand{\Tb}{\bar{T}_b}

\title[HI and cosmological constraints]
{HI and cosmological constraints from intensity mapping, optical, and CMB surveys}
\author[ Pourtsidou, Bacon, Crittenden]{ Alkistis Pourtsidou, David Bacon, Robert Crittenden  \\ 
Institute of Cosmology \& Gravitation, University of Portsmouth, Burnaby Road, Portsmouth, PO1 3FX, United Kingdom 
}

\begin{document}

\maketitle

\begin{abstract}
We forecast constraints on neutral hydrogen (HI) and cosmological parameters using near-term intensity mapping surveys with instruments such as BINGO, MeerKAT, and the SKA, and Stage III and IV optical galaxy surveys. If foregrounds and systematic effects can be controlled -- a problem which becomes much easier in cross-correlation -- these surveys will provide exquisite measurements of the HI density and bias, as well as measurements of the growth of structure, the angular diameter distance, and the Hubble rate, over a wide range of redshift. We also investigate the possibility of detecting the late time ISW effect using the Planck satellite and forthcoming intensity mapping surveys, finding that a large sky survey with Phase 1 of the SKA can achieve a near optimal detection. 
\end{abstract}


\section{Introduction}

According to the standard cosmological model, dark energy is responsible for the current acceleration of the Universe \citep{Riess:1998cb, Perlmutter:1998np}. This is supported by a wealth of high precision cosmological data, such as measurements of supernovae (e.g. \citep{Kowalski:2008ez, Lampeitl:2009jq, Betoule:2014frx}), the Cosmic Microwave Background (CMB) \citep{Komatsu:2010fb, Ade:2015xua}, and baryon acoustic oscillations (e.g. \citep{Blake:2011en, Anderson:2013zyy, Delubac:2014aqe, Aubourg:2014yra}. In order to study dark energy's possible evolution and ultimately uncover its nature, large scale galaxy surveys like the Dark Energy Survey (DES)\footnote{www.darkenergysurvey.org}  and the Euclid satellite\footnote{www.euclid-ec.org} are being commissioned. At the same time, a state-of-the-art radio telescope, the Square Kilometre Array (SKA)\footnote{www.ska.ac.za}, is being built; it aims to answer questions about the formation and evolution of the first galaxies, cosmic magnetism, dark matter, dark energy, and gravity.  

In recent years, there has been a significant amount of work on the prospects of precision cosmology studies with the SKA and other radio instruments using the neutral hydrogen (HI) intensity mapping (IM) technique (see, for example, \citet{Bull:2014rha,Santos:2015gra,Pourtsidou:2015mia}). Intensity mapping \citep{Battye:2004re,Chang:2007xk,Loeb:2008hg,Mao:2008ug,Peterson:2009ka,Seo:2009fq,Ansari:2011bv,Battye:2012tg,Switzer:2013ewa} is a novel technique that uses HI as a dark matter tracer in order to map the 3D large-scale structure (LSS) of the Universe. It measures the intensity of the redshifted 21cm line, hence it does not need to detect galaxies but treats the 21cm sky as a diffuse background. This means that intensity mapping surveys can scan large areas of the sky very quickly, and perform high precision clustering measurements. 

In order for the aforementioned optical galaxy and intensity mapping surveys to reach the unprecedented statistical precision required ($< 1\%$ for large sky surveys with Euclid or SKA), systematic effects must be controlled and removed. In the case of HI intensity mapping, for example, we have the presence of large galactic and extragalactic foregrounds. These are orders of magnitude bigger than the signal we want to recover, but they are expected to have a smooth frequency dependence so that they can be removed \citep{LiuTegmark,  Chang:2010jp, Wolz:2013wna, Alonso15, Bigot-Sazy:2015tot, Olivari:2015tka, Switzer:2015ria, Wolz:2015lwa}. One way to accelerate and improve HI detection and intensity mapping cosmology is to cross-correlate the 21cm maps with optical galaxy surveys, in which case a large part of the noise and systematic effects are expected to be uncorrelated and drop out. In fact, the only HI detections we currently have using intensity mapping come from cross-correlating the 21cm maps from the Green Bank Telescope (GBT) with optical galaxy surveys \citep{Chang:2010jp, Masui:2012zc, Switzer:2013ewa}. We note that such cross-correlations will benefit optical galaxy surveys as well, since they also suffer from large-scale systematic effects. In general, combining surveys is expected to yield measurements more robust and precise than any individual experiment.

In this paper, we will demonstrate how near-term 21cm intensity mapping and optical galaxy surveys can be used to constrain HI and cosmological parameters. Our work builds upon a previous paper \citep{Pourtsidou:2015mia}, in which we explored the potential of using intensity mapping and optical galaxy surveys to detect HI clustering and weak gravitational lensing and showed that high precision measurements can be performed. In this work, we will first forecast HI and cosmological parameter constraints using 21cm IM auto-correlation clustering measurements, and then move on to cross-correlations with optical galaxy surveys. We will also show how the late Integrated Sachs-Wolfe (ISW) effect can be detected by cross-correlating 21cm IM maps with the CMB maps from the Planck satellite mission \citep{Ade:2015xua}. As we have already mentioned, the performance of the 21cm surveys depends on how well foregrounds can be subtracted, as well as on calibration and other problems on very large scales. We have not accounted for these issues in our auto-correlation forecasts, so they should be seen as a ``best-case" scenario. Since foreground removal and control of systematics should be much easier in cross-correlation -- which is the main focus of this paper --  we believe that our cross-correlation forecasts are more realistic and robust. Similar conclusions were reached recently in \citet{Wolz:2015ckn}, where the authors studied the cross-correlation of intensity mapping observations with optical galaxy surveys at $z \approx 0.9$ implementing different star formation models. It was shown that the systematic error contribution to the auto-power spectrum measurements can be very large resulting in significant disagreement between theoretical modelling and observations, while cross-correlations perform much better and the modelling is in good agreement with the data from the GBT cross-correlated with WiggleZ galaxies \citep{Masui:2012zc}. 

The plan of the paper is as follows: In Section \ref{sec:surveys} we describe the range of HI intensity mapping and optical galaxy surveys we are going to use for our forecasts. In Section \ref{sec:constraintsHI} we consider auto-correlation HI clustering measurements and employ the Fisher matrix formalism to derive constraints for HI and cosmological parameters. In Section \ref{sec:crosscorr} we extend the formalism to include cross-correlations with optical galaxy surveys. Finally, in Section \ref{sec:ISW} we describe how we can detect the late-time integrated Sachs-Wolfe (ISW) effect in cross-correlation using HI intensity mapping and the Planck satellite. We conclude in Section \ref{sec:conclusions}.

\section{The surveys}
\label{sec:surveys}

\subsection{Intensity Mapping Surveys}

We consider a range of HI intensity mapping surveys, focussing on the SKA and its precursor MeerKAT. As discussed in detail in previous works \citep{Bull:2014rha}, the telescope arrays can be used in auto-correlation (``single-dish") mode. This allows the instrument to probe large scales, which is not possible when the array functions in the conventional interferometric mode.
We also consider a single-dish experiment, BINGO\footnote{http://www.jb.man.ac.uk/research/BINGO/}, which can be thought of as an excellent pathfinder in order to test and exploit the intensity mapping technique, and establish the possible issues that we will face with more advanced IM surveys using MeerKAT and the SKA. 

\subsection*{BINGO}

BINGO \citep{Battye:2012tg} is a single dish intensity mapping project, whose main goal is to detect HI using the 21cm intensity mapping technique and measure the Baryon Acoustic Oscillation scale in the redshift range $z=0.12-0.48$ (corresponding to a frequency range of $300 \, {\rm MHz}$). It will scan approximately $2000 \, {\rm deg}^2$ in 1 year total observation time with an angular resolution $\theta_B \simeq 40 \, {\rm arcmin}$ and $N_{\rm beams} \sim 50$ feeds. Its system temperature is expected to be $T_{\rm sys} \simeq 50 \, {\rm K}$. 

The survey's noise properties have been described in detail in \citet{Battye:2012tg}, but we will summarise them here for completeness. The frequency resolution of IM surveys is very good, so we can ignore the instrument response function in the radial direction and only consider the response due to the finite angular resolution:
\be
W^2(k)={\rm exp}\left[-k_\perp^2r(z)^2\left(\frac{\theta_{\rm B}}{\sqrt{8{\rm ln}2}}\right)^2\right],
\label{eq:response}
\ee where $\mathbf{k}_\perp$ is the transverse wavevector, $r(z)$ is the comoving radial distance at redshift $z$ and $\theta_{\rm B} \sim \lambda/D_{\rm dish}$ the beam full width at half maximum (FWHM) of a single dish with diameter $D_{\rm dish}$ at wavelength $\lambda$. 
Considering a redshift bin with limits $z_{\rm min}$ and $z_{\rm max}$, the survey volume will be given by
\be
V_{\rm sur}=\Omega_{\rm tot} \int_{z_{\rm min}}^{z_{\rm max}} dz \frac{dV}{dz d\Omega} =  \Omega_{\rm tot} \int_{z_{\rm min}}^{z_{\rm max}} dz \frac{c r(z)^2}{H(z)},
\label{eq:volume}
\ee and $ \Omega_{\rm tot} = A_{\rm sky}$, the sky area the survey scans. The pixel volume $V_{\rm pix}$ is also calculated from Eq.~(\ref{eq:volume}), but with 
\be
 \Omega_{\rm pix} \simeq 1.13\theta^2_B
\ee assuming a Gaussian beam, and the corresponding pixel $z$-limits corresponding to the channel width $\Delta f$, which is $\sim 1 \, {\rm MHz}$ for BINGO. Finally, the pixel noise $\sigma_{\rm pix}$ is given by
\be
\sigma_{\rm pix}=\frac{T_{\rm sys}}{\sqrt{\Delta f \, t_{\rm total}(\Omega_{\rm pix}/\Omega_{\rm tot})N_{\rm dishes} N_{\rm beams}}},
\ee with $N_{\rm dishes}$ the number of dishes ($N_{\rm dishes} = 1$ for BINGO) and $t_{\rm total}$ the total observing time. 
In intensity mapping experiments, the dominant noise contribution comes from the thermal noise of the instrument. The noise power spectrum is then given by
\be
\label{eq:PNsd}
P^{\rm N}(k) = \sigma^2_{\rm pix}V_{\rm pix}W^{-2}(k) \, .
\ee

\subsection*{MeerKAT}

MeerKAT\footnote{http://www.ska.ac.za/science-engineering/meerkat/} is a 64-dish SKA precursor on the planned site of SKA1-MID, with 20 dishes already in place. 
A detailed description of the noise properties of the MeerKAT dishes can be found in \citet{Pourtsidou:2015mia}.
MeerKLASS, a large sky intensity mapping survey with MeerKAT, has been proposed\footnote{The MeerKLASS survey -- Mario Santos, MeerKAT Science: On the Pathway to the SKA, 25-27 May, 2016, Stellenbosch, South Africa.}. It will scan a few thousand square degrees on the sky (we will take $A_{\rm sky} = 4000 \, {\rm deg}^2$ here) in approximately $4000$ hours total observation time. 
There are two bands available, namely the L band $0<z<0.58$ ($900<f<1420$ MHz), and the UHF band $0.4<z<1.45$ ($580<f<1000$ MHz); we will show constraints for both bands, taking $z=0.6$ as the lower $z$ limit for the UHF band. The array will operate in single dish mode, which means that the auto-correlation signal from the dishes is considered and very large scales can be probed if enough sky is scanned. This means that its noise power spectrum $P^{\rm N}$ is given by Equation~(\ref{eq:PNsd}) with the appropriate parameters for the MeerKAT dishes \citep{Pourtsidou:2015mia}.

\subsection*{SKA1} 

We consider Phase 1 of the SKA\_MID instrument, 
consisting of 130 dishes with 15 m diameter according to the recently updated specifications (re-baselining) \citep{McPherson15}. The redshift range is $0.35<z<3$ (Band 1) and the expected system temperature is $T_{\rm sys}=25 \, {\rm K}$. The potential of SKA1 to perform an HI intensity mapping survey over a broad range of frequencies and a large fraction of the sky ($f_{\rm sky}\sim 0.7$) has been presented in \citet{Santos:2015gra}. Such a survey can deliver precision measurements of baryon acoustic oscillations, redshift space distortions (RSDs), and weak gravitational lensing, and constrain the spatial curvature of the Universe, primordial non-Gaussianity, and the sum of neutrino masses.

\subsection{Optical Galaxy Surveys}

Current and future state-of-the-art optical galaxy surveys aim to investigate the nature of dark energy by scanning large areas of the sky and combining multiple cosmological probes, for example Supernovae, Baryonic Acoustic Oscillations, galaxy clusters and weak gravitational lensing. Here we will consider a Stage III photometric optical galaxy survey, similar to the ongoing Dark Energy Survey (DES)\footnote{http://www.darkenergysurvey.org/}, and a future, Stage IV spectroscopic survey, similar to the one that will be performed by the Euclid satellite\footnote{http://www.euclid-ec.org/}. 

In galaxy surveys the noise power spectrum is dominated by shot noise, which is due to the discreteness of the galaxies sampling the underlying matter density and is given by
\be
P^{\rm shot} = \frac{1}{\bar{n}} = \frac{1}{(N_g/V_{\rm sur})}\, ,
\ee with $N_g$ the number of galaxies within the redshift bin under consideration with comoving volume $V_{\rm sur}$.

\subsection*{Stage III}

We consider a Stage III photometric optical galaxy survey with $A_{\rm sky}=5000 \, {\rm deg}^2$, number density of galaxies $n_g = 10 \, {\rm arcmin}^{-2}$, and redshift range $0<z<2$ with median redshift $z_0=0.7$. 
In order to calculate the shot noise contribution for this survey we will use the above specifications and a redshift distribution of the form \citep{Asorey:2012rd}
\be
\frac{dn}{dz} \propto z^\alpha \, {\rm exp}[-(z/z_0)^\eta]\, , 
\ee with $\alpha=2, \eta=3/2$. The photometric redshift error is taken to be $\sigma_z \simeq 0.05(1+z)$.
In order to model the galaxy bias on large scales we will use $b_{\rm g}=\sqrt{1+z}$ \citep{Rassat:2008ja}. 

\subsection*{Stage IV}

We consider a Stage IV spectroscopic optical galaxy survey in the redshift range $0.7<z<2$ detecting tens of millions of galaxies in a sky area $A_{\rm sky}=15000 \, {\rm deg}^2$.  In our forecasts for such a survey we will use the number density of galaxies $\bar{n}$ given in \citet{Majerotto:2015bra}, where the predicted redshift distribution has been split into 14 bins with $\Delta z=0.1$.

We will now proceed to forecast HI and cosmological constraints from intensity mapping auto-correlation measurements, as well as from cross-correlating the 21cm maps with the optical galaxies.

\section{Constraints from HI intensity mapping alone}
\label{sec:constraintsHI}

\subsection{HI Power Spectrum}

The mean 21cm emission brightness temperature is given by (see \citet{Battye:2012tg} for a detailed derivation)
\be
\Tb(z) = 180 \Omega_{\rm HI}(z)h\frac{(1+z)^2}{H(z)/H_0} \, {\rm mK} \, ,
\ee where $\Omega_{\rm HI}$ is the HI density, $H(z)$ the Hubble parameter as a function of redshift $z$, and $H_0\equiv 100h$ its value today.
Neglecting for the moment the effect of redshift space distortions, the signal (S) HI power spectrum can be written as
\be
P^{\rm S} \equiv P^{\rm HI}(k,z)=\Tb^2 b^2_{\rm HI}P(k,z) \, ,
\ee where $P$ is the matter power spectrum, and $b_{\rm HI}$ the HI bias.
We can further write
\be
P(k,z)=D^2(z)P(k,z=0) \, ,
\ee where $D(z)$ is the growth factor.

The Fisher matrix for a set of parameters $\{p\}$ is then given by \citep{Tegmark:1997rp, Seo:2007ns}
\be
F_{\rm ij} = \frac{1}{2}\left[C^{-1}\frac{\partial C}{\partial p_i}C^{-1}\frac{\partial C}{\partial p_j}\right] 
+ \frac{\partial \mu^T}{\partial p_i}C^{-1}\frac{\partial \mu^T}{\partial p_j}    \, ,
\ee where $C$ is the covariance matrix and $\mu$ the model for the tested parameters.
From the Cramer-Rao inequality, we know that the best errors we can achieve are given by
\be
\Delta p_i \geq (F_{\rm ii})^{-1/2} \, .
\ee
Following \citet{Feldman:1993ky}, we can write 
$\mathbf{\mu}_n \approx P^{\rm S}(k_n)$ in a thin Fourier shell of radius $k_n$ and 
\be
C_{mn} \approx \frac{2}{V_nV_{\rm sur}}[P^{\rm S}(k_n)+1/\bar{n}]^2\delta_{\rm mn} \, ,
\ee where $V_n \equiv 4\pi k^2_n dk_n/(2\pi)^3$ is the volume element and $dk_n$ the width of the shell. Then we can define the ``effective volume'' as
\be
V_{\rm eff}(k_n) \equiv \left[\frac{\bar{n}P^{\rm S}(k_n)}{1+\bar{n}P^{\rm S}(k_n)}\right]^2 V_{\rm sur} \, , 
\ee with $\bar{n}$ number density of galaxies and $V_{\rm sur}$ the survey volume.
For thick shells that contain many uncorrelated modes the Fisher Matrix can be written as \citep{Tegmark:1997rp}
\be
F_{\rm ij} \approx \frac{1}{4\pi^2}\int^{k_{\rm max}}_{k_{\rm min}} k^2dk \; [\partial_i {\rm ln}P^{\rm S} \partial_j {\rm ln}P^{\rm S} ] V_{\rm eff} \, ,
\label{eq:Fish1}
\ee where we have also replaced the sum by an integral. 
For the case of IM surveys we can write a similar formula \citep{Bull:2014rha, Pourtsidou:2015mia}
with
$
V_{\rm eff} = V_{\rm sur} \left(\frac{P^{\rm S}}{P^{\rm S}+P^{\rm N}}\right)^2 ,
$ where $P^{\rm N}$ is the noise power spectrum defined in Equation~(\ref{eq:PNsd}). 
In the following we will assume that the bias $b_{\rm HI}$ depends only on the redshift $z$, i.e. that it is scale-independent. This assumption is appropriate only for large (linear) scales, so we will impose a non-linear cutoff at $k_{\rm max} \simeq 0.14 (1+z)^{2/3} \, {\rm Mpc}^{-1}$ \citep{Smith:2002dz}. Hence, we will also ignore the small scale velocity dispersion effects (``fingers of god"). The largest scale the survey can probe corresponds to a wavevector $k_{\rm min} \simeq 2\pi/V^{1/3}$.
Inverting the Fisher matrix we get the covariance matrix that gives us the forecasted constraints on the chosen parameters set. 

As a first approach we assume a flat $\Lambda$CDM expansion history and keep all cosmological parameters fixed to the Planck 2015 cosmology \citep{Ade:2015xua}. Then the only unknown in the HI power spectrum ($P^{\rm HI}$) is the prefactor $\Omega_{\rm HI}b_{\rm HI}$.
Measuring this quantity is very important for future IM surveys, as the magnitude of the power spectrum affects the signal-to-noise ratio. Pinning down the evolution of the HI density and bias will help us forecast and optimise the scientific output of large sky IM surveys with the SKA.

Knowing how the HI density $\Omega_{\rm HI}$ evolves with redshift is very important not only for cosmology, but also for galaxy evolution and star formation history studies. At very low redshifts $z < 0.3$, HI galaxy surveys like HIPASS can measure $\Omega_{\rm HI}$ \citep{Zwaan:2005cz}, while at high redshifts $z > 2$ damped Ly$\alpha$ systems can be used (see \citet{Crighton:2015pza} and references therein).   

\begin{figure}
\includegraphics[scale=0.6]{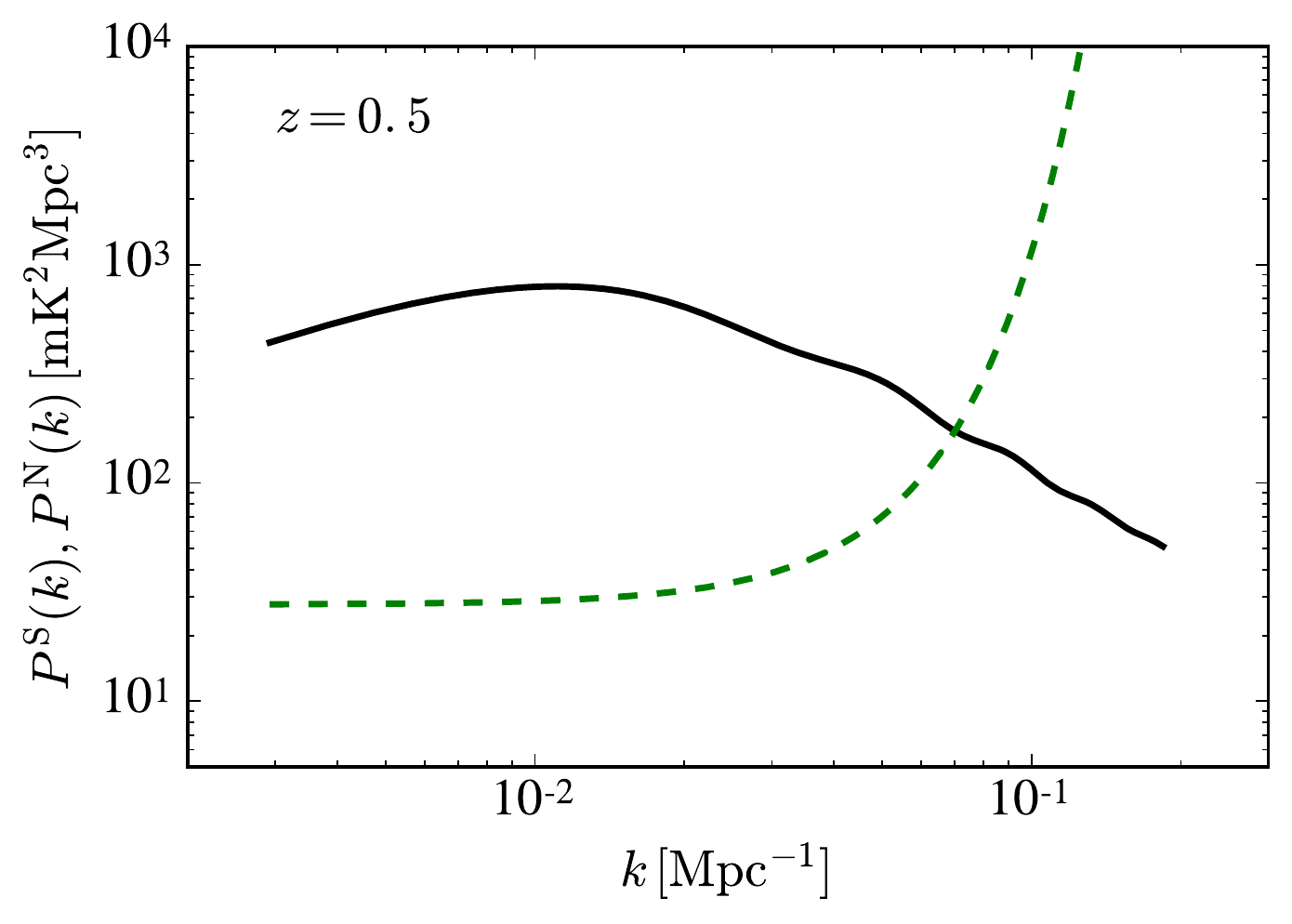}
\caption{The signal (solid black line) and thermal noise (dashed green line) power spectra at $z=0.5$ for our fiducial cosmology and the chosen MeerKAT IM survey parameters (see main text for details). }
\label{fig:pspn}
\end{figure}

We will now employ the formalism described above to forecast constraints that can be achieved by the MeerKAT IM survey. 
We calculate the expected constraints on $\Omega_{\rm HI}(z)b_{\rm HI}(z)$ across the whole redshift range covered by the L and UHF bands, but we note that the actual survey is going to use only one of the bands. We consider a series of -- independent -- redshift bins across $0<z<1.45$, with width $\Delta z = 0.1$, and use the central redshift of each bin for our calculations. 
For our fiducial models of the HI density and bias we use the fits from \citet{Bull:2014rha}. 
As an example of the level of signal-to-noise ratio from such a survey, we plot $P^{\rm S} \equiv P^{\rm HI}$ and $P^{\rm N}$ for the bin with central redshift $z=0.5$ in Figure~\ref{fig:pspn}, setting $k_\perp \sim k$ in Eq.~(\ref{eq:response}) for simplicity. We see that the noise diverges as we reach the limits set by the beam resolution --- this means that, even if we did not impose a linear cut-off for the $k_{\rm max}$ value in the Fisher matrix, we would not get significant improvement (except at very low $z$) as the signal-to-noise ratio for non-linear scales is dropping fast.

Our results for the fractional uncertainties for $\Omega_{\rm HI}b_{\rm HI}$ are presented in Table~\ref{tab:omHIbHI}.
We see that sub-percent measurements can be achieved at $z<0.6$ (L band) and $1-4\%$ measurements for $z>0.6$ (UHF band). These are at least one order of magnitude better than the currently available constraints from galaxy surveys, intensity mapping, and damped Lyman-$\alpha$ observations (see Table 2 in \citet{Padmanabhan:2014zma}). Repeating the same procedure for BINGO -- considering one bin with central redshift $z=0.3$ and $\Delta z \simeq 0.35 $ -- we find $\delta(\Omega_{\rm HI}b_{\rm HI})/(\Omega_{\rm HI}b_{\rm HI}) \simeq 0.003$. 

\begin{table}
\begin{center}
\caption{\label{tab:omHIbHI} Forecasted fractional uncertainties on $\Omega_{\rm HI}b_{\rm HI}$ and $\Omega_{\rm HI}$  at various redshifts, assuming the MeerKAT IM survey specifications described in the main text. For the $\Omega_{\rm HI}$ constraints we utilise the full HI power spectrum with RSDs.}  
\begin{tabular}{ccc}
\hline
$z$ & $\delta(\Omega_{\rm HI}b_{\rm HI})/(\Omega_{\rm HI}b_{\rm HI})$ & $\delta \Omega_{\rm HI}/\Omega_{\rm HI}$\\
\hline
{\bf L band} \\
0.1&0.010&0.06\\
0.2&0.005&0.03\\
0.3&0.005&0.03\\
0.4&0.007&0.03\\
0.5&0.009&0.04\\
{\bf UHF band} \\
0.6&0.011&0.04\\
0.7&0.013&0.04\\
0.8&0.015&0.05\\
0.9&0.018&0.06\\
1.0&0.022&0.07\\
1.1&0.026&0.08\\
1.2&0.030&0.09\\
1.3&0.036&0.10\\
1.4&0.042&0.12\\
\hline
\end{tabular}
\end{center}
\end{table}

\subsection{Redshift Space Distortions}

The full HI signal power spectrum in redshift space can be written as 
\be
P^{\rm S} \equiv P^{\rm HI}(k,z;\mu)=\Tb^2 b^2_{\rm HI} [1+\beta_{\rm HI}(z)\mu^2]^2P(k,z) \, ,
\ee where $P$ is the matter power spectrum, $b_{\rm HI}$ the HI bias, $\mu = \hat{k} \cdot \hat{z}$ and $\beta_{\rm HI}$ the redshift space distortion parameter equal to $f/b_{\rm HI}$ in linear theory, where $f \equiv d{\rm ln}D/d{\rm ln}a$ is the linear growth rate with the scale factor $a=1/(1+z)$. 

In \citet{Masui:2010mp}, the authors showed that redshift space distortions measurements can break the degeneracy between $\Omega_{\rm HI}$ and $b_{\rm HI}$.
Assuming a flat $\Lambda$CDM expansion history, we have $f\simeq\Omega(z)^{0.545}$, and from that we can calculate $D(z)$. Assuming the Planck 2015 cosmology we parametrise $P^{\rm HI}$ by two redshift dependent parameters, namely $\beta$ and the combination $ b^2_{\rm HI}\Omega^2_{\rm HI}$. As discussed in \citet{Masui:2010mp}, the amplitude parameter $(b_{\rm HI}\Omega_{\rm HI})^2$ will be more precisely measured than $\beta$ and does not contribute much to the $\Omega_{\rm HI}$ uncertainty; measuring $\beta$ we measure $b_{\rm HI}$ and, subsequently, we get a measurement of $\Omega_{\rm HI}$:
\be
\frac{\delta \Omega_{\rm HI}}{\Omega_{\rm HI}} \approx \frac{\delta \beta}{\beta} \, .
\ee 

\begin{figure}
\includegraphics[scale=0.6]{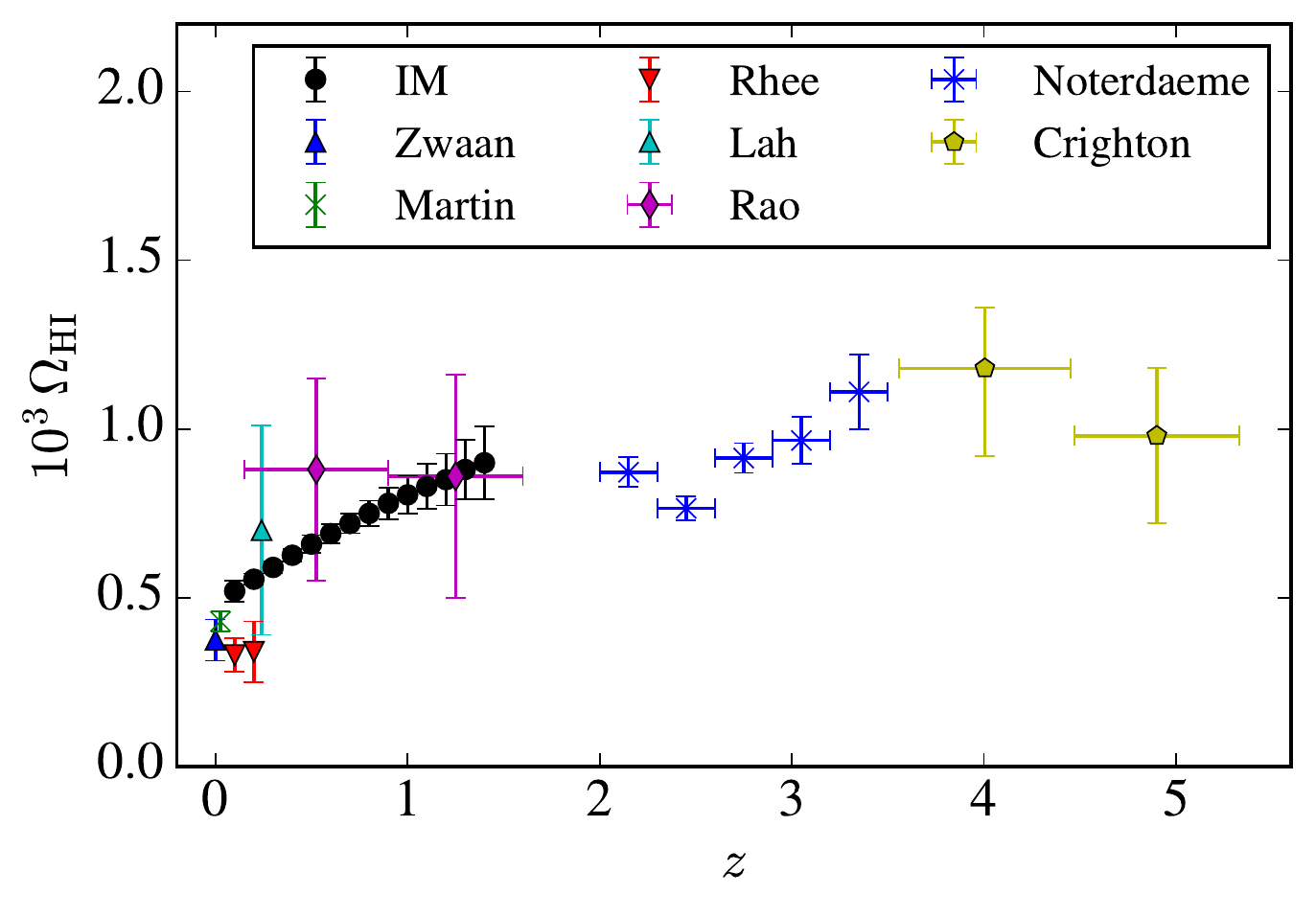}
\caption{ Measurements of the HI density, $\Omega_{\rm HI}$, across redshift. We include the results of \citep{Zwaan:2005cz,Rao:2005ab,Martin:2010ij,Noterdaeme:2012gi,Rhee:2013fma} (see \citet{Crighton:2015pza} for a complete list), and add our forecasted data for an intensity mapping (IM) survey with MeerKAT.}
\label{fig:omHI}
\end{figure}

The Fisher matrix is now given by \citep{Tegmark:1997rp, Seo:2007ns}
\be
F_{\rm ij}=\frac{1}{8\pi^2}\int^{1}_{-1} d\mu \int^{k_{\rm max}}_{k_{\rm min}} k^2dk \; [\partial_i {\rm ln}P^{\rm S} \partial_j {\rm ln}P^{\rm S} ] V_{\rm eff} \, .
\label{eq:Fish2}
\ee 
Our results for the fractional uncertainties for $\Omega_{\rm HI}$ are presented in Table~\ref{tab:omHIbHI}. Note that for the noise modelling we use the full anisotropic response function, Eq.~(\ref{eq:response}), with $k_\perp = k\sqrt{1-\mu^2}$. Again, the results are much better than the currently available (see Table 2 in \citet{Padmanabhan:2014zma}). In order to demonstrate the potential impact of such measurements, we add our forecasts using the MeerKAT IM survey to the current data coming from HI galaxy surveys and DLAs in Figure~(\ref{fig:omHI}). As we can see, the IM measurements can fill the gap between the low and high redshift observations. The combined measurements can be used to constrain halo models for cosmological neutral hydrogen \citep{Padmanabhan:2016odj}, and to understand the clustering properties of HI by measuring the HI bias (note, for example, that the HI bias and DLAs bias are different, and that intensity mapping measures the sum of all HI) \citep{Castorina:2016bfm,Sarkar:2016lvb}. It is important to note that these measurements are key in order to discriminate between models of how to assign HI to the dark matter haloes $M_{\rm HI}(M)$. For example, the simulations of \citet{Villaescusa-Navarro:2015zaa} show that AGN feedback reduces $M_{\rm HI}$ and hence suppresses the whole curve, while \citep{Padmanabhan:2016odj} suggest a slope effect on the high mass end of the HI-halo mass relation.

Repeating the above procedure for BINGO (one redshift bin with $0.12<z<0.48$) we find $\frac{\delta \Omega_{\rm HI}}{\Omega_{\rm HI}} \simeq 0.02$.

We will now modify our approach and assume that the mean temperature $\bar{T}(z)$ has been measured. 
One way to achieve this is by measuring the smooth part of the HI signal, but global measurements are instrumentally very challenging. However, if one can take advantage of the spectral variation in the brightness temperature, the HI signal can be separated from the foregrounds. This should be doable in Epoch of Reionization measurements, but calibrating the data to the required level is much harder in the low redshifts we consider in this work.  
Nevertheless, we will consider the HI density known, either from RSD measurements like we described above, or from cross-correlation measurements, HI galaxy surveys, and damped Lyman-$\alpha$ systems in combination with results from simulations. Alternatively, one could include $\bar{T}_b$ in the RSD functions in the Fisher matrix analysis.

A comprehensive study forecasting cosmological constraints for large sky intensity mapping surveys has been presented in \citet{Bull:2014rha}. Here we will first perform a similar analysis in order to see how well a large sky IM survey with MeerKAT can constrain the growth of structure; this will be followed by forecasts using cross-correlations with optical galaxy clustering data. We consider the two approaches (three if we include the measurements coming from optical galaxy surveys alone) highly complementary and synergistic. 

Our forecasting method builds upon previous work on geometrical and growth constraints from galaxy redshift surveys like Euclid (see, for example, \citet{Samushia:2010ki, Majerotto:2015bra}).    
Our chosen parameters are the Hubble rate $H(z)$, the angular diameter distance $D_{\rm A}(z)$, $f\sigma_8(z)$, and $b_{\rm HI}\sigma_8(z)$, which are considered to be independent in each redshift bin. The first two encode important information about the expansion and geometry of the Universe, through their dependance on the matter density $\Omega_m$, the Hubble parameter $h$, the spatial curvature, and the dark energy density $\Omega_{\rm DE}$, while $f\sigma_8(z)$ encodes information about the growth of structure. We choose to work with these parameters because they are model-independent. If we want to forecast the constraints on a specific cosmological model, for example one that can be parameterised by a dark energy equation of state $w=w_0+w_a z/(1+z)$ \citep{Chevallier:2000qy, Linder:2002et} and a growth index $\gamma$ \citep{Linder:2007hg}, we can use a Fisher transformation matrix to move from the old parameters to the new parameters. 

The expression of the power spectrum (divided by $\bar{T}^2_b)$ with respect to our chosen parameters
is \citep{Seo:2007ns}
\be
\frac{P^{\rm HI}(k_{\rm f},\mu_{\rm f},z)}{\bar{T}^2_b}= \frac{D_{\rm A}(z)^2_{\rm f}H(z)}{D_{\rm A}(z)^2H(z)_{\rm f}}
 (b_{\rm HI}\sigma_8(z)+f\sigma_8(z)\mu^2)^2\frac{P(k,z)}{\sigma^2_8(z)} \, ,
\ee where the subscript ``f" refers to the fiducial (``reference") cosmology. Note that the derivatives in the corresponding Fisher matrix are all analytic, except $\partial P(k,z)/\partial k$ that enters the derivatives with respect to $D_{\rm A}$ and $H$. That is because 
$k_{f\parallel}=k_{\parallel}H(z)_{\rm f}/H(z)$, $k_{f\perp}=k_{\perp}D_{\rm A}(z)/D_{\rm A}(z)$.

We perform the Fisher matrix calculations by considering the natural logarithm of the aforementioned parameters. 
The marginalised constraints on $\{{\rm ln}f\sigma_8, {\rm ln}D_{\rm A}, {\rm ln}H\}$ for the MeerKAT survey configuration -- i.e. the forecasted fractional uncertainties on $\{f\sigma_8, D_{\rm A}, H\}$ -- are presented in Table~\ref{tab:fsig8}. Here, we also show the covariance matrix corresponding to the full parameter set $\{{\rm ln}f\sigma_8, {\rm ln}b_{\rm HI}\sigma_8, {\rm ln}D_{\rm A}, {\rm ln}H\}$ for the redshift bin centred at $z=0.5$:
$$
\begin{bmatrix} 
0.0019 & -0.00008 & 0.0010 & -0.00004 \\
 & 0.0007 & 0.0004 & -0.0007 \\
 & & 0.0009 & -0.0004 \\
 & & & \;\;\; 0.0007  
\end{bmatrix}
$$
From Table~\ref{tab:fsig8} we can see that our chosen $4000 \, {\rm deg}^2$ survey with MeerKAT can achieve around $4\%$ constraints on $f\sigma_8$ in the L band, i.e. at redshifts where dark energy dominates. Performing the survey in the UHF band worsens the constraints because of the lower signal-to-noise ratio of the measurements, but it is a unique opportunity to probe a relatively unexplored cosmological epoch tomographically. 
The effect of the anisotropic response function of a single-dish IM survey (with excellent resolution along the line-of-sight but large beam effects along the transverse direction that increase with redshift) is evident -- for example, the $D_{\rm A}$ error increases quickly along the UHF band, while the $H$ error is ``flatter".
In any case, such a survey is able to provide growth and expansion history constraints for the first time using intensity mapping, and complement and compete with state-of-the-art optical galaxy surveys (for a detailed investigation of the possibilities for precision late-time cosmology with 21cm intensity mapping using a variety of radio instruments, see \citet{Bull:2014rha}). As we have already stated, the above forecasts are an optimistic scenario, since we have assumed that foregrounds have been removed and we have ignored instrumental systematics and calibration issues at large scales.

A toy model to assess how good the foreground removal has to be in order for IM surveys like the one we are assuming here to yield useful cosmological constraints has been studied in \citet{Bull:2014rha}). 
The contribution from major foregrounds like extragalactic point sources, free-free emission, and galactic synchrotron has been parametrised in terms of amplitude and index parameters, and a foreground removal method is assumed. The ``cleaned'' maps will then have some residual contamination, which is parametrised by an overall amplitude $\epsilon^2$, with $\epsilon=1$ corresponding to no removal. There is also a related parameter, namely a minimum radial wavenumber $k_{\rm FG}$; the larger that is, the smaller $\epsilon$ is (more foreground removal), but obviously there is a tradeoff with loss of cosmological signal. In \citet{Bull:2014rha} it is found that $\epsilon < 10^{-5}$ is needed in order to extract the cosmological information, while $\epsilon = 10^{-6}$ can be achieved using current foreground subtraction techniques.

We will now move on to consider constraints coming from cross-correlating intensity mapping maps with optical galaxy data. This is expected to attenuate the above issues.

\begin{table}
\begin{center}
\caption{\label{tab:fsig8} Forecasted fractional uncertainties on $\{f\sigma_8, D_{\rm A}, H\}$ at each redshift bin, assuming the MeerKAT IM survey specifications described in the main text.}  
\begin{tabular}{cccc}
\hline
$z$ & $\delta(f\sigma_8)/(f\sigma_8)$ & $\delta D_{\rm A}/D_{\rm A}$ & $\delta H/H$\\
\hline
{\bf L band} \\
0.1&0.08&0.06&0.07\\
0.2&0.04&0.03&0.04\\
0.3&0.03&0.02&0.03\\
0.4&0.04&0.03&0.03\\
0.5&0.04&0.03&0.03\\
{\bf UHF band} \\
0.6&0.05&0.03&0.03\\
0.7&0.06&0.04&0.03\\
0.8&0.07&0.05&0.03\\
0.9&0.08&0.06&0.04\\
1.0&0.10&0.07&0.04\\
1.1&0.11&0.08&0.05\\
1.2&0.13&0.09&0.05\\
1.3&0.15&0.10&0.06\\
1.4&0.17&0.12&0.06\\
\hline
\end{tabular}
\end{center}
\end{table}

\section{Constraints from cross-correlations}
\label{sec:crosscorr}

In this Section we will investigate what can be achieved from cross-correlating HI intensity mapping with optical galaxy surveys. Before we set out the formalism for cross-correlations, it is worth taking a step back and discussing optical galaxy surveys alone.
For an optical galaxy survey, the large scale signal $P^{\rm gg}(k,z)$ is given by
\be
P^{\rm gg}(k,z) = b^2_{\rm g}P(k,z) \, ,
\ee with $b_{\rm g}$ the galaxy bias, which we model as linear and deterministic. If we include redshift space distortions, this becomes
\be
P^{\rm gg}(k, \mu, z) = b^2_{\rm g}[1+\beta(z)\mu^2]^2P(k,z) \, ,
\ee with $\beta = f /b_{\rm g}$.
The effective survey volume entering the Fisher matrix calculation for the optical galaxy survey auto-correlation measurements can be written as \citep{Seo:2007ns}
\be
V_{\rm eff} = V_{\rm sur} \left(\frac{P^{\rm gg}(k,\mu,z){\rm exp}[-(k\mu\sigma_r)^2]}{P^{\rm gg}(k,\mu, z){\rm exp}[-(k\mu\sigma_r)^2]+1/\bar{n}}\right)^2 \, ,
\label{eq:Veffgg}
\ee with $\sigma_r=c\sigma_z/H(z)$, where $\sigma_z$ is the redshift error. For a spectroscopic survey like Euclid, this is very small ($\sim 0.001$) and can effectively be ignored, but it will be much larger for a photometric survey ($\sim 0.1$). This means that radial information is lost and the power is strongly suppressed along the line of sight modes $k_\parallel$ \citep{Blake:2004tr, Seo:2007ns}. For this reason, it is more difficult for photometric surveys to measure observables like $H(z)$ or redshift space distortions. 

Let us now set up the formalism for the 21cm IM -- optical cross-correlations. 
In \citet{Masui:2012zc} the cross-correlation of the 21cm maps acquired at the Green Bank Telescope and the galaxies of the WiggleZ spectroscopic survey were used to constrain the quantity $\Omega_{\rm HI}b_{\rm HI}r$ -- with $r$ a correlation coefficient accounting for the possible stochasticity in the galaxy and HI tracers -- at $z\sim 0.8$ with a statistical fractional error $\sim 16\%$.  The cross-correlation power spectrum can be written as \citep{Chang:2010jp, Masui:2012zc}
\be
P^{\rm HI,g}(k,z) = \bar{T}_b b_{\rm HI}b_{\rm g}r P(k,z) \, ,
\ee where we have neglected redshift space distortions (we will include them in the next Section). Assuming a flat $\Lambda$CDM expansion history keeping all cosmological parameters fixed to the best-fit Planck cosmology, and using the measurement of $b_g$ by the galaxy survey, as in \citet{Masui:2012zc}, the only unknown is the prefactor $\Omega_{\rm HI}b_{\rm HI}r$. 

Here we will first consider a forthcoming Stage IV spectroscopic survey (similar to Euclid), which is assumed to have a $500 \, {\rm deg}^2$ overlap with MeerKAT. This is a conservative assumption since we do not yet know which part of the sky will be scanned by the MeerKAT IM survey.
As we have already stated, the Euclid spectroscopic redshift range is $0.7<z<2$, which overlaps with the UHF band for MeerKAT.
Our fiducial model is as before and we set the fiducial stochastic correlation coefficient $r=1$. 

The Fisher matrix in this case will be given by Equation~(\ref{eq:Fish1}) with 
\be
V_{\rm eff} = V_{\rm sur}\frac{P^{\rm HI,g}(k,z)^2}{P^{\rm HI,g}(k,z)^2+P^{\rm HI,tot}(k,z)P^{\rm g,tot}(k,z)} \, ,
\label{eq:Veffcross}
\ee
with $P^{\rm HI,tot} = P^{\rm HI}+P^{\rm N}$ the total (signal+noise) HI power spectrum, and $P^{\rm g,tot} = P^{\rm gg}+1/\bar{n}$. We take $\bar{n}$ to be similar to the mean number density (in each redshift bin) used in \citet{Majerotto:2015bra}. The constraints from a Stage IV spectroscopic survey and MeerKAT with $500 \, {\rm deg}^2$ overlap on the sky are summarised in Table~\ref{tab:omHIbHIr}. We can see that even with a ``small" (compared to the sky area of the individual surveys) sky overlap we can measure $\Omega_{\rm HI}b_{\rm HI}r$ across redshift with quite good precision --- allowing $4000 \, {\rm deg}^2$ overlap we get a factor of $\sim 3$ improvement in the constraints. Note that for an ``intermediate" spectroscopic survey with the same sky and redshift overlap, but with $100$ times smaller number density of galaxies in each bin, we get constraints that are a factor of $\sim 2$ worse. 

As we have already mentioned, the abundance and bias properties of HI are of great importance for both astrophysics (for example determining the HI luminosity function) and cosmology. The capabilities and potential science output of future HI intensity mapping surveys depend on the amplitude of the signal $\Omega_{\rm HI}b_{\rm HI}$. Better measurements of $b_{\rm HI}$ and $r$ are needed, but we can assume that -- especially on large scales -- the value of $r$ is scale independent and consistent with unity \citep{Khandai:2010hs,Wolz:2015ckn}. Therefore, cross-correlation studies can put a lower limit on $\Omega_{\rm HI}b_{\rm HI}$ \citep{Masui:2012zc}.

\begin{table}
\begin{center}
\caption{\label{tab:omHIbHIr} Forecasted fractional uncertainties on $\Omega_{\rm HI}b_{\rm HI}r$ at various redshifts, assuming MeerKAT and a Stage IV spectroscopic survey. The sky overlap is taken to be $500 \, {\rm deg}^2$ and we have reduced the total observation time (with respect to the full $4000 \, {\rm deg}^2$ survey) accordingly.}  
\begin{tabular}{cc}
\hline
$z$ & $\delta(\Omega_{\rm HI}b_{\rm HI}r)/(\Omega_{\rm HI}b_{\rm HI}r)$\\
\hline
{\bf UHF band} \\
0.6&--\\
0.7&0.06\\
0.8&0.07\\
0.9&0.08\\
1.0&0.10\\
1.1&0.12\\
1.2&0.14\\
1.3&0.16\\
1.4&0.18\\
\hline
\end{tabular}
\end{center}
\end{table}

\subsection{Redshift Space Distortions}

Including redshift space distortions in the IM-optical galaxy cross-correlation formalism (see \citet{White:2008jy} for the case of multiple galaxy populations), we can write 
\be
P^{\rm HI,g}(k,\mu,z)  =  \bar{T}_b b_{\rm HI} b_{\rm g} [1+\beta_{\rm HI}\mu^2][1+\beta_{\rm g}\mu^2] P(k,z) \, .
\ee
 Note that we have not included stochastic coefficients ($r$ factors), i.e. we have implicitly assumed that they are scale-independent and equal to unity on large scales. 

We are first going to consider the combination of MeerKAT and a Stage III photometric optical galaxy survey, with a $4000 \, {\rm deg}^2$ overlap. We are going to incorporate the photometric redshift error $\sigma_z \simeq 0.05(1+z)$ by replacing  \citep{Seo:2007ns}
\bea \nonumber 
&& P^{\rm HI,g} \rightarrow P^{\rm HI,g} \, {\rm exp}[-k^2\mu^2(c\sigma_z)^2/H(z)^2/2] \\ \nonumber
&& P^{\rm gg} \rightarrow P^{\rm gg} \, {\rm exp}[-k^2\mu^2(c\sigma_z)^2/H(z)^2]
\eea in the calculation of $V_{\rm eff}$. 
The fact that radial information is lost means that $\beta_{\rm HI}$ is not going to be measured well, so our primary focus here is measuring the amplitude
$\Omega_{\rm HI}b_{\rm HI}$.

We will assume a $\Lambda$CDM expansion history and keep all cosmological parameters fixed to the Planck 2015 cosmology, in order to examine what constraints can be obtained on a minimal set of parameters. Furthermore, we will suppose that the galaxy bias $b_{\rm g}$ is known (i.e. measured from the galaxy survey). We therefore parameterise $P^{\rm HI,g}$ using two parameters, namely  ($\Omega_{\rm HI}b_{\rm HI}$, $\beta_{\rm HI}$), and use the aforementioned fiducial models for $\Omega_{\rm HI}$ and $b_{\rm HI}$ in our forecasts. We present the marginalised constraints for $\Omega_{\rm HI}b_{\rm HI}$ in Table~\ref{tab:omHIbHIcross}. Even if $\beta$ is poorly measured, we can constrain $\Omega_{\rm HI}b_{\rm HI}$ at a $\sim 5\%$ level across a wide range of redshift. This is still much better than the currently available constraints on $\Omega_{\rm HI}b_{\rm HI}$.
We note that while the power spectrum approach in this section is convenient for predicting parameter constraints, in practice an analysis of photometric redshift survey data is likely to use the angular (2D) galaxy clustering power spectrum \citep{Nock:2010gr, Ross:2011zza}. In this way, RSD information can be recovered from cross-correlations between different redshift bins \citep{Asorey:2013una}. 

It is worth commenting on how the above constraints change if we assume a much smaller, spectroscopic-like redshift error, which we take to be one order of magnitude smaller than the DES-like error, i.e. we consider $\sigma_z = 0.005$. We find that there is a factor of $\sim 2$ improvement in the $\Omega_{\rm HI}b_{\rm HI}$ constraints across the L-band, and a smaller improvement at higher redshifts. The biggest effect of course comes in measuring $\beta$, as radial information is now retained: we get a $\sim 10\%$ fractional error from $z=0.2$ to $z=0.8$, which corresponds to a measurement of the HI density since
 the degeneracy between $\Omega_{\rm HI}$ and $b_{\rm HI}$ is broken (as we described in the previous Section). 

\begin{table}
\begin{center}
\caption{\label{tab:omHIbHIcross} Forecasted fractional uncertainties on $\Omega_{\rm HI}b_{\rm HI}$, assuming MeerKAT and a Stage III photometric survey. The sky overlap is taken to be $4000 \, {\rm deg}^2$.}  
\begin{tabular}{cc}
\hline
$z$ & $\delta(\Omega_{\rm HI}b_{\rm HI})/(\Omega_{\rm HI}b_{\rm HI})$\\
\hline
{\bf L band} \\
0.1&0.09\\
0.2&0.05\\
0.3&0.04\\
0.4&0.04\\
0.5&0.04\\
{\bf UHF band} \\
0.6&0.05\\
0.7&0.05\\
0.8&0.06\\
0.9&0.07\\
1.0&0.07\\
1.1&0.09\\
1.2&0.10\\
1.3&0.11\\
1.4&0.12\\
\hline
\end{tabular}
\end{center}
\end{table}

Modifying our approach as before, we now suppose that $\bar{T}_b$ is known and we focus on measuring the growth of structure across redshift. We will consider the combination of a Stage IV spectroscopic survey with an intensity mapping survey performed using SKA1, assuming a $7000 \, {\rm deg}^2$ overlap and $4000$ hours total observation time. We write the cross-correlation power spectrum as 
\bea \nonumber
P^{\rm HI,g}(k_{\rm f},\mu_{\rm f},z) &=& \frac{D_{\rm A}(z)^2_{\rm f}H(z)}{D_{\rm A}(z)^2H(z)_{\rm f}}
 (b_{\rm HI}\sigma_8(z)+f\sigma_8(z)\mu^2) \\
&&\times (b_{\rm g}\sigma_8(z)+f\sigma_8(z)\mu^2)\frac{P(k,z)}{\sigma^2_8(z)} \, ,
\eea and we suppose that $b_{\rm g}\sigma_8$ has been measured from the galaxy survey, so we keep it fixed. We will therefore use the parameter set $\{{\rm ln}f\sigma_8, {\rm ln}b_{\rm HI}\sigma_8,{\rm ln}D_{\rm A},{\rm ln}H\}$ in our Fisher matrix. We summarise the (marginalised) forecasted constraints in Table~\ref{tab:fsig8cross}. 
We note that we have checked our Fisher matrix code results are in agreement with the ones presented in \citet{Majerotto:2015bra} when we consider a Stage IV galaxy survey only and constrain $\{{\rm ln}f\sigma_8, {\rm ln}b_{\rm g}\sigma_8,{\rm ln}D_{\rm A},{\rm ln}H\}$ using the $P^{\rm gg}(k,\mu,z)$ power spectrum.

We see that we get very good constraints on the growth of structure, the Hubble parameter and the angular diameter distance across a wide range of redshifts corresponding to the era where dark energy is starting to become important ($0.7<z<1.4$). Once again, we stress that the above forecasts are expected to be more trustworthy and robust than the auto-correlation ones presented in Table~\ref{tab:fsig8}, due to the alleviation of systematic effects.
\begin{table}
\begin{center}
\caption{\label{tab:fsig8cross} Forecasted fractional uncertainties on $\{f\sigma_8, D_{\rm A}, H\}$ at each redshift bin, assuming the SKA1 IM and Stage IV spectroscopic survey specifications described in the main text.}  
\begin{tabular}{cccc}
\hline
$z$ & $\delta(f\sigma_8)/(f\sigma_8)$ & $\delta D_{\rm A}/D_{\rm A}$ & $\delta H/H$\\
\hline
{\bf UHF band} \\
0.6&--&--&--\\
0.7&0.04&0.03&0.02\\
0.8&0.05&0.03&0.02\\
0.9&0.05&0.03&0.03\\
1.0&0.06&0.04&0.03\\
1.1&0.07&0.04&0.03\\
1.2&0.08&0.05&0.03\\
1.3&0.10&0.06&0.03\\
1.4&0.11&0.06&0.04\\
\hline
\end{tabular}
\end{center}
\end{table}

\section{Prospects for ISW detection with intensity mapping surveys}
\label{sec:ISW}

In this section we will consider the possibility of detecting the ISW effect in cross-correlation using neutral hydrogen (HI) as a tracer of the underlying LSS. The late ISW effect \citep{Sachs:1967er} arises from the time variation of the gravitational potential and can be detected by cross-correlating the CMB with a low redshift tracer of the matter distribution (see, for example, \citet{Boughn:2004zm, Fosalba:2003ge, Giannantonio:2008zi, Giannantonio:2012aa}). Detection of the ISW effect in a flat Universe provides direct evidence for dark energy.

\subsection{The ISW effect in cross-correlation}

Let us assume some tracer of matter (tr) --- this can be e.g. galaxies (g) or neutral hydrogen (HI). The tracer density contrast we observe in a direction $\hat{\bn}_1$ will be
\be
\delta_{\rm tr}(\hat{\bn}_1) = \int b_{\rm tr}(z) W(z) \delta_m(\hat{\bn}_1,z)dz \, ,
\ee
where $W(z)$ is the selection function of the survey, $b_{\rm tr}(z)$ the tracer bias and $\delta_m$ the matter density perturbations. The observed tracer density will be correlated with the ISW temperature fluctuation in a direction $\hat{\bn}_2$, which is
\be
\frac{\Delta T}{T}(\hat{\bn}_2) = -2\int {\rm e}^{-\tau(z)}\frac{d\Phi}{dz}(\hat{\bn}_2,z)  dz \, ,
\ee where $\Phi$ is the gravitational potential and ${\rm e}^{-\tau(z)}$ is the visibility function of the photons.

Having a CMB map and a tracer (galaxy or HI) survey, the auto- and cross-correlation power spectra will be given by
\bea
&&C^{\rm T-tr}_\ell = 4\pi \int \frac{dk}{k}\Delta^2(k) {\cal I}^{\rm ISW}_\ell (k) {\cal I}^{\rm tr}_\ell(k) \, , \\
&&C^{\rm tr-tr}_\ell = 4\pi \int \frac{dk}{k}\Delta^2(k) [{\cal I}^{\rm tr}_\ell (k)]^2 \, ,
\eea where
$\Delta(k)$ is the scale invariant matter power spectrum $\Delta^2(k) \equiv 4\pi k^3 P(k)/(2\pi)^3$, and the two integrands are respectively
\bea
&{\cal I}^{\rm ISW}_\ell (k) = -2\int {\rm e}^{-\tau(z)} \frac{d\Phi_k}{dz}j_\ell[kr(z)]dz \, , \\
&{\cal I}^{\rm tr}_\ell (k) = \int b_{\rm tr}(z) W(z)D(z)j_\ell[kr(z)]dz \, ,
\eea where $\Phi_k$, $\delta_m(k,z)$ are the Fourier components of the gravitational potential and matter perturbations, and $j_\ell$ are the spherical Bessel functions. 

\subsection{Using HI as a tracer}

Letting ${\rm tr} = {\rm HI}$ the density contrast can be written in terms of the 21cm temperature fluctuations, $\delta_{\rm HI}(\hat{\bn}_1) \equiv \delta T_{\rm b}/\bar{T}_{\rm b}$ -- note that we will assume thin enough bins ($\Delta z = 0.1$) so that we can consider $\bar{T}_{\rm b}$ constant within the bin, equal with its value at the central bin redshift $z_c$.
Using the Limber approximation \citep{Limber:1954zz, LoVerde:2008re} we find

\bea \nonumber
C^{\rm T-HI}_\ell &=& - \frac{3\Omega_{m,0}(H_0/c)^3}{(\ell +1/2)^2} \int dz W(z) E(z) b_{\rm HI}(z) \\ 
&\times& P_0((\ell+1/2)/r(z))D^2(z)(f(z)-1) \, , \\ \nonumber
C^{\rm HI-HI}_\ell &=& \frac{H_0}{c}\int dz E(z) W^2(z) P_0((\ell+1/2)/r(z)) \\
&\times& b^2_{\rm HI}(z)D^2(z)/r^2(z),
\eea where $E(z) \equiv H(z)/H_0$. Note that if we set ${\rm HI} \rightarrow {\rm g}$ we recover the literature results for the galaxy case (see, for example, \citet{Afshordi:2004kz, Giannantonio:2008zi, Francis:2009ps}).

\begin{figure}
\includegraphics[scale=0.6]{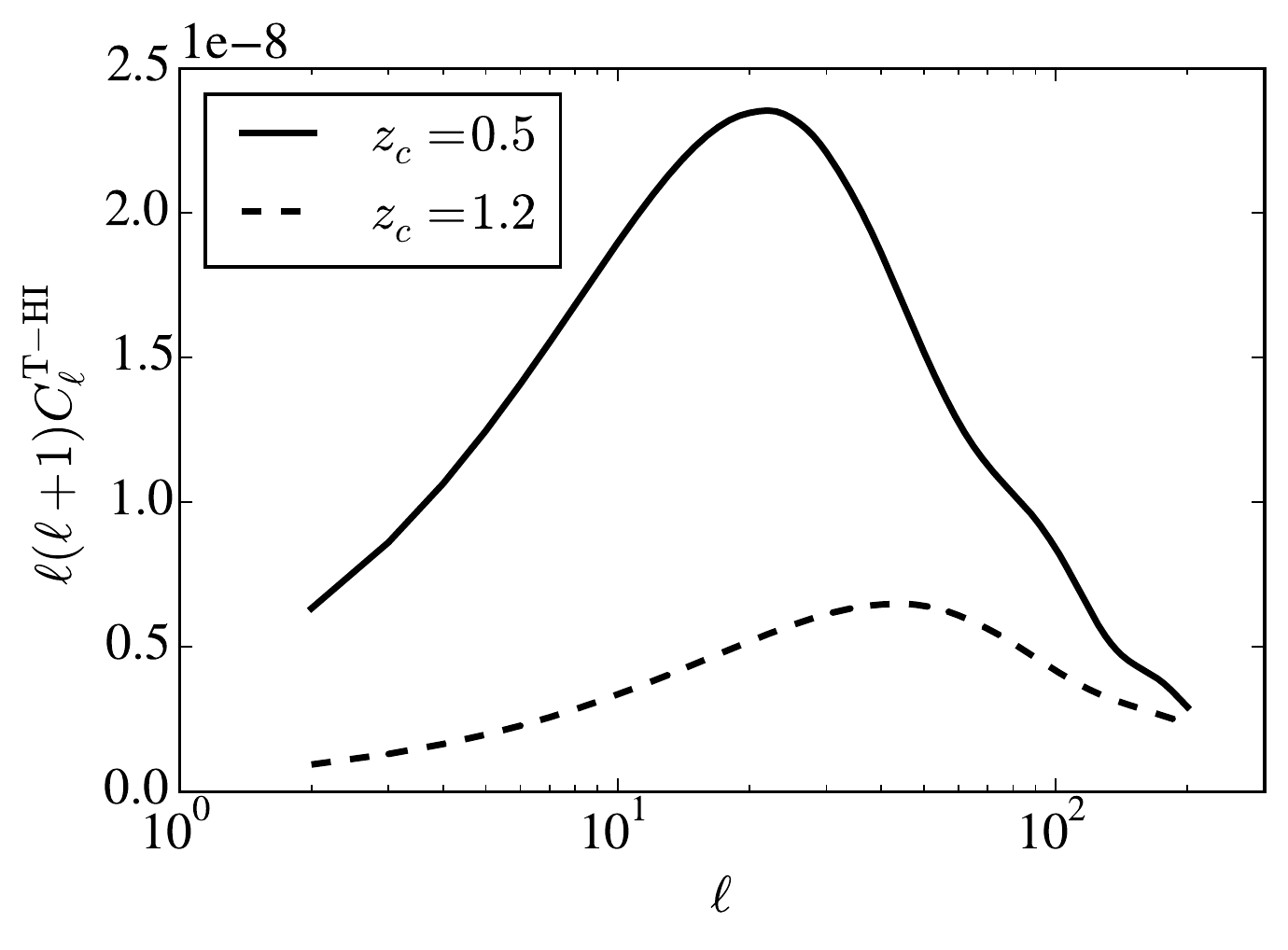}
\caption{The angular spectrum of the correlation of the Planck CMB temperature at $z_c=0.5$ (solid black line) and $z_c=1.2$ (dashed black line) for our fiducial cosmology. The bin width is $\Delta z =0.1$.}
\label{fig:CTHI}
\end{figure}

The characteristics of the cross-correlation signal have been studied extensively (see \citet{Afshordi:2004kz} for a detailed analysis), but we will reproduce the main results here for completeness. Our fiducial cosmology is the Planck $\Lambda$CDM best-fit model \citep{Ade:2015xua}.
 In Figure~\ref{fig:CTHI} we show the angular power spectrum $C^{\rm T-HI}$ for two redshifts $z_c=0.5$ and $z_c=1.2$ taking a $\Delta z = 0.1$ bin in both cases. 
The ISW cross-correlation is distributed over a wide redshift range $0<z<2$, but the main contribution comes from $z\sim 0.4$. Also note that the signal lies on scales larger than a degree, $\ell < 200$. 

\subsection{Signal-to-Noise forecasts}

Having the formalism set up, we can now investigate what can be achieved with forthcoming intensity mapping surveys. 
Using HI as a tracer and an IM survey, the signal-to-noise ratio (SNR) will be given by
\be
\left(\frac{S}{N}\right)^2 = f_{\rm sky} \sum^{\ell_{\rm max}}_{\ell=\ell_{\rm min}} (2\ell+1) \frac{[C_\ell^{\rm T-HI}]^2}{(C_\ell^{\rm HI-HI}+N^{\rm HI}_\ell)C_\ell^{\rm TT}+[C_\ell^{\rm T-HI}]^2} \, ,
\label{eq:snr}
\ee where $N^{\rm HI}_\ell$ is the instrumental noise of the telescope performing the IM survey, and $f_{\rm sky}$ the (overlapping) fraction of the sky the CMB and IM surveys scan. We have ignored the instrumental noise of the Planck satellite (which would add to $C_\ell^{\rm TT}$) because it is negligible, especially on large scales.

Let us first assume that we have the  ``perfect survey", i.e. set $N^{\rm HI}_\ell=0$ and $f_{\rm sky}=1$. This will give us the maximum SNR we can achieve. We assume that the survey covers the redshift range $0<z<3$ and we will split it in bins with $\Delta z=0.1$, and take $W(z)=1/\Delta z$, a top-hat function. For these forecasts we will set $b_{\rm HI}=1$ for simplicity.

The total signal-to-noise ratio is calculated by applying Equation~(\ref{eq:snr}) at each independent redshift bin and then adding the $(S/N)^2$.
For our assumed cosmology we find $(S/N)_{\rm tot} \sim 6.5$, in agreement with previous results assuming a perfect galaxy survey \citep{Crittenden:1995ak, Afshordi:2004kz}.
Below we present our results for the various intensity mapping surveys we have considered in this paper; the results are also summarised in Table~\ref{tab:ISW}. 
Note that the noise $N_\ell^{\rm HI}$ in Eq.~(\ref{eq:snr}) is given by \citep{Battye:2012tg, Pourtsidou:2015mia}
\be
N_\ell^{\rm HI} = \Omega_{\rm pix}(T_{\rm sys}^2/2 B t_{\rm obs}) {\rm exp}[\ell(\ell+1)(\theta_B/\sqrt{8{\rm ln}2})^2]/\bar{T}_b^2 \, ,
\label{eq:Nelldish}
\ee with $B$ the frequency bandwidth of observation. 

\subsubsection*{BINGO}

Using the formalism outlined above and the BINGO telescope parameters we find $S/N < 1$. This is not surprising, since the BINGO redshift range ($0.12<z<0.48$) and sky coverage ($2000 \, {\rm deg}^2$) are not large enough --- even if we ignore the instrumental noise $N_\ell^{\rm HI}$, the signal-to-noise ratio is below unity. 

\subsubsection*{MeerKAT}

Our best chance for detecting the ISW effect with a $4000 \, {\rm deg}^2$ MeerKAT survey (in cross-correlation with Planck) is performing the IM survey in the UHF band, which corresponds to the redshift range $0.4<z<1.45$. We find $S/N=1.5$ and, again, neglecting the instrumental noise $N_\ell^{\rm HI}$ only slightly increases the significance of the detection. 

\subsubsection*{SKA-Mid, Phases 1 and 2}

In single dish mode, SKA1-Mid can perform an all-sky intensity mapping survey with useful area $f_{\rm sky}=0.7$. We are going to consider Band 1 (B1), with $0.35<z<3.06$ in our forecasts, and an observation time of $5000$ hours.  We find $S/N=4.6$. Considering an SKA2-like survey -- with an order of magnitude lower thermal (instrumental) noise -- we find  $S/N=4.9$, which is practically the maximum signal-to-noise-ratio that can be achieved by a survey with $f_{\rm sky}=0.7$ and redshift range $0.35<z<3.06$. 
This is competitive with the forecasted performance of next generation large scale structure optical galaxy surveys like Euclid and LSST\footnote{www.lsst.org} \citep{Douspis:2008xv, Refregier:2010ss}. Finally, we note that the possibility of detecting the ISW effect using high-redshift ($z\sim 30$) 21-cm maps cross-correlated with galaxies has been studied in \citet{Raccanelli:2015lca}, where it was shown that a good detection requires very advanced radio instruments (for example lunar interferometers).

\begin{table}
\begin{center}
\caption{\label{tab:ISW} Forecasts of the signal-to-noise ratio (SNR) for detecting the late time ISW effect using Planck and IM surveys.}  
\begin{tabular}{cccccc}
\hline
IM Survey &$z$ range&$f_{\rm sky}$&SNR\\
\hline
Perfect survey&0--3&1&6.3\\
BINGO &0.12--0.48&0.05&0.7\\
MeerKAT UHF-band &0.4--1.45&0.1&1.5\\
SKA1-MID Band 1 &0.35--3.06&0.7&4.6\\
SKA2-like Band 1 &0.35--3.06&0.7&4.9\\
\hline
\end{tabular}
\end{center}
\end{table}

\section{Conclusions}
\label{sec:conclusions}

In this paper, we have shown how ongoing and near-term intensity mapping and optical galaxy surveys can be used to constrain the Universe's expansion history and the growth of cosmic structure, as well as the evolution of HI. We forecasted HI and cosmological constraints for a range of HI surveys and their cross-correlation with galaxies, with special emphasis on the performance of the SKA and its pathfinder MeerKAT. 

Our auto-correlation forecasts show that precision measurements of HI and cosmological parameters can be performed already with a $\sim 4000 \, {\rm deg}^2$ intensity mapping survey with MeerKAT, across a wide range of redshift. Constraining the HI evolution and bias tomographically before the SKA comes online is very important for maximising its scientific output, and probing the HI abundance and evolution across cosmic time is of key importance for astrophysics and cosmology alike. 
Our cross-correlation forecasts establish that measurements of exquisite precision can be made combining 21cm intensity maps and optical galaxies, with the extra advantage of alleviating major issues like systematic effects and foreground contaminants, which are relevant for one type of survey but not for the other. Finally, we showed that a large sky survey with Phase 1 of the SKA combined with the Planck temperature maps can detect the ISW effect at a level competitive with state-of-the-art Stage IV optical galaxy surveys. 

We believe that the results of this paper provide strong motivation for exploring further the possibilities of cross-correlations between 21cm intensity mapping, optical galaxies, and the CMB. 

\section*{Acknowledgments}
AP's research is supported by a Dennis Sciama Fellowship at the University of Portsmouth. We acknowledge use of the \texttt{CAMB} code \citep{camb} and \texttt{Astropy} \citep{Robitaille:2013mpa}. We would like to thank Roy Maartens, Andrej Obuljen, and Mario Santos for useful discussions. Fisher matrix codes used in this work are available from \url{https://github.com/Alkistis/IM-Fish}.

\bibliographystyle{mn2e}
\bibliography{references_HIconstr}

\end{document}